\documentclass[twocolumn,aps,superscriptaddress]{revtex4}
\usepackage{palatino}
\usepackage[latin1]{inputenc}
\usepackage{epsf}
\usepackage{amsmath,amssymb}
\usepackage{latexsym}
\usepackage{calc}
\usepackage{color}
\usepackage{shadow}
\usepackage{epsfig}

\def\mol{_{\sss mol}}

\newcommand{\ben}{\begin{equation}}
\newcommand{\een}{\end{equation}}
\newcommand{\bea}{\begin{eqnarray}}
\newcommand{\eea}{\end{eqnarray}}

\def\sss{\scriptscriptstyle\rm}
\def\x{_{\sss X}}
\def\c{_{\sss C}}
\def\s{_{\sss S}}
\def\xc{_{\sss XC}}

\def\H{_{\sss H}}

\def\ext{_{\rm ext}}

\def\br{{\bf r}}

\def\half{\frac{1}{2}}
\def\ee{_{\rm ee}}

\def\n{n}

\begin{document}

\title{Dynamics of Charge-Transfer Processes with Time-Dependent Density Functional Theory}
%: Nonlocality in Space and Time}

\author{J. I. Fuks}
\affiliation{Nano-Bio Spectroscopy group, Dpto.~F\'isica de Materiales, Universidad del Pa\'is Vasco, Centro de
F\'isica de Materiales CSIC-UPV/EHU-MPC and DIPC, Av.~Tolosa 72, E-20018 San 
Sebasti\'an, Spain}
\author{P. Elliott}
\affiliation{Max-Planck-Institut f\"{u}r Mikrostrukturphysik, Weinberg 2, 06120 Halle (Saale), Germany}
\author{A. Rubio}
\affiliation{Nano-Bio Spectroscopy group, Dpto.~F\'isica de Materiales, Universidad del Pa\'is Vasco, Centro de
F\'isica de Materiales CSIC-UPV/EHU-MPC and DIPC, Av.~Tolosa 72, E-20018 San 
Sebasti\'an, Spain}
\affiliation{Fritz-Haber-Institut der Max-Planck-Gesellschaft, Faradayweg 4-6,
D-14195 Berlin, Germany}
\author{N. T. Maitra}
\affiliation{Department of Physics and Astronomy, Hunter College and the Graduate Center of the City University of New York, 695 Park Avenue, New York, New York 10065, USA}
\date{\today}

\begin{abstract}
We show that whenever an electron transfers between closed-shell molecular
fragments, the exact correlation potential of
time-dependent density functional theory develops a step and
peak structure in the bonding region. This structure has a
density-dependence that is non-local both in space and time, that even
the exact adiabatic ground-state exchange-correlation functional fails to
capture it.  For charge-transfer between
open-shell fragments, an initial step and peak vanish as the
charge-transfer state is reached.  The inability of usual approximations to develop these structures leads to inaccurate charge-transfer dynamics. This is
illustrated by the complete lack of Rabi oscillations in
the dipole moment under conditions of resonant charge-transfer for an exactly-solvable model system. The results transcend the model and are applicable to more realistic molecular complexes.
%Neepa: since being in an excited state is needed for the argument of why a step develops, should we mention here that the CT is to an excited state? i.e. we don't really know or prove anything about whether a CT-step develops when an electron transfers across the system to some arbitrary time-evolving step, right?
%{\it develop the topic of the last sentence and the two below indicated issues in a later version.}
%\newline
%{\bf should we specify excited state?}
%\newline
%{\bf should we mention dynamical step? gets too long?}
\end{abstract}
\maketitle

Charge-transfer (CT) dynamics play a critical role in many processes of
interest in physics, chemistry, and biochemistry, from photochemistry to photosynthesis, solar cell design and biological functionality. 
%capture of solar energy leading to a useful end state is an area of
%active interest which often requires simulating electron transfer
The quantum mechanical treatment of such systems calls for methods
that can treat electron correlations and dynamics efficiently for relatively
large systems. Time-dependent density functional theory
(TDDFT)~\cite{RG84,newTDDFTbook} is the leading candidate today, and has achieved an unprecedented balance between accuracy and efficiency in calculations of electronic spectra  \cite{newTDDFTbook,PCCPissue}.
%Neepa -- I just added a citation to the whole issue, is that ok or would it be better to give specific articles?
CT excitation energies over medium to large distances are, however, notoriously underestimated by the usual exchange-correlation (xc) functionals, and recent years have witnessed intense development of many methods to treat it~\cite{A09,SKB09,BLS10,HG12}.
There is recent optimism for obtaining accurate CT excitations
between closed-shell fragments ~\cite{SKB09,BLS10}, but
no functional approximation
developed so far works for 
CT between open-shell
fragments~\cite{M05c,MT06,FRM11}. Here standard approximations predict even an unphysical ground-state with fractional occupation in the
dissociation limit.  For open-shell fragments the exact ground-state
correlation potential has step and peak structures \cite{TMM09,HTR09}, while the exact
xc kernel has strong frequency-dependence and
diverges as a function of the fragment separation; lack of these
features in the xc-approximation is responsible for their poor
predictions. 

In contrast to linear
response phenomena, the description of
photoinduced processes 
%such as excited state reactivity,
%photoisomerisation, photofragmentation, electron and proton transfer
%and electron-hole recombination, 
generally requires a complete
electron transfer from one state to another, or from different regions
of space. This is the case in photovoltaic materials (organic, inorganic, and hybrids), photocatalysis, biomolecules in solvents, reactions at the interface between different materials, nanoscale conductance devices (see e.g. Refs~\cite{PDP09,Oviedo,Teoh,Nyuen,Moore,Meng} and references therein). 
These processes are clearly nonlinear and require a
non-perturbative {\it time-resolved} study of electron dynamics rather
than a simple calculation of their excitation spectrum.  
TDDFT is increasingly used, often within an Ehrenfest or surface-hopping scheme  to handle coupled electron-ion motion~\cite{newTDDFTbook,PDP09,Oviedo,Teoy,Nyuen,Moore,Meng}. In the TDDFT scheme, 
a one-body time-dependent Kohn-Sham (KS) potential is used to evolve a set of non-interacting KS electrons, 
reproducing the exact one-body density of the true interacting system, 
 from which all properties of the interacting system may be exactly
extracted. 
In practise, approximations are required for the xc potential,
$v\xc[n;\Psi_0,\Phi_0](\br,t)$, a functional of the one-body density
$n$, the initial interacting state $\Psi_0$ and the initial KS state
$\Phi_0$. Almost all calculations today use an adiabatic
approximation, that inserts the instantaneous density into a
ground-state xc approximation, $v\xc^{adia}[n;\Psi_0,\Phi_0](\br,t) =
v\xc^{g.s.}[n(t)](\br,t)$, neglecting the dependence of $v\xc$ on the
past history and initial states~\cite{newTDDFTbook}. Further,
the exact $v\xc$ has in general a non-local dependence on space
%, the ground-state approximation most often has a spatially local or semi-local dependence on the density
\cite{fnote1}.

A critical question is: Are the available functionals suitable for
modeling the CT processes mentioned earlier? 
%This is  
% How would these approximations perform for the time-resolved
% CT process?  
In this paper, we show that when an electron
  transfers at long range from a ground- to an excited
 CT- state, a time-dependent step and
 peak are  generic and essential features of the exact xc potential. 
%These require the correlation potential to have spatially non-local and non-adiabatic dependence on the density, features lacking in the approximations used today.
When the donor and acceptor are both closed shells, the initial xc potential has no step nor peak, but a step and peak
 structure in the bond midpoint region builds up
 over time.
 Although in the initial stages of the CT dynamics the usual approximations may perform well,
%be good approximations to the exact xc potential,
 they are increasingly worse as time evolves, leading to completely wrong long-time dynamics.
%(see Fig.~\ref{f:R7}).
% The reverse is true
% for the case where 
On the other hand, when the donor and acceptor are both open-shell
species, an initial step and peak structure wanes.  Thus these
time-dependent steps and peaks that are difficult to capture in
functional approximations, play a significant role in CT {\it even
  between neutral closed-shell fragments, unlike in the calculation of
  excitation energies}.  Further, we show that although an adiabatic
approximation to the xc potential may yield a step structure, the step
will, at best, be of the wrong size.  Accompanying the step and peak
associated with charge transfer there is also a dynamical
step~\cite{EFRM12}, that depends on how the CT is achieved.  The exact
$v\xc$ thus has a complicated non-local space- and time-dependence
that adiabatic functionals fail to capture, with severe consequences
for time-resolved CT. Although our results are demonstrated for two
electrons, we expect they can be generalized to real molecular
systems, as many cases of CT dominantly involve two valence
electrons. The other electrons act as a general buffer that introduces
some additional dynamical screening that can change the net size of the
step and peak but not their presence.

%mainly are involved two-valence electrons and that the rest acts a a general buffer that introduces some additional dynamical screening, tthat cna change the size of the step and peak but not their presence that is a genaral feature of CT processes.

%, and that require ultranon-local spatial-dependence on the density.
%The  consequence for 
%of the lack of step features in approximate functionals on 
%the time-resolved dynamics is severe, as we will show for resonant CT processes.
%\ben
%i\frac{\partial}{\partial t}\phi_j(\br,t) = \left(-\frac{\nabla^2}{2} + v\s(\br,t) \right) \phi_j(\br,t)
%\een
%where
%\ben
%v\s(\br,t) = v\ext(\br,t) + v\H[n(t)](\br,t) + v\xc[n,\Psi_0,\Phi_0](\br,t)
%\een
%where $v\ext(\br,t)$ is external potential including electron-nuclei interaction and external fields, $v\H(\br,t)$ is the Hartree potential, and $v\xc(\br,t)$ is the xc potential. The Hartree potential depends instantaneously on the density, while the xc potential has functional dependence on the entire history of the density, the initial interacting wavefunction, and the initial KS wavefunction. Thus we speak of the xc functional having memory. 

\begin{figure}[ht]
\begin{center}
\includegraphics[width=0.475\textwidth,height=0.22\textwidth,clip]{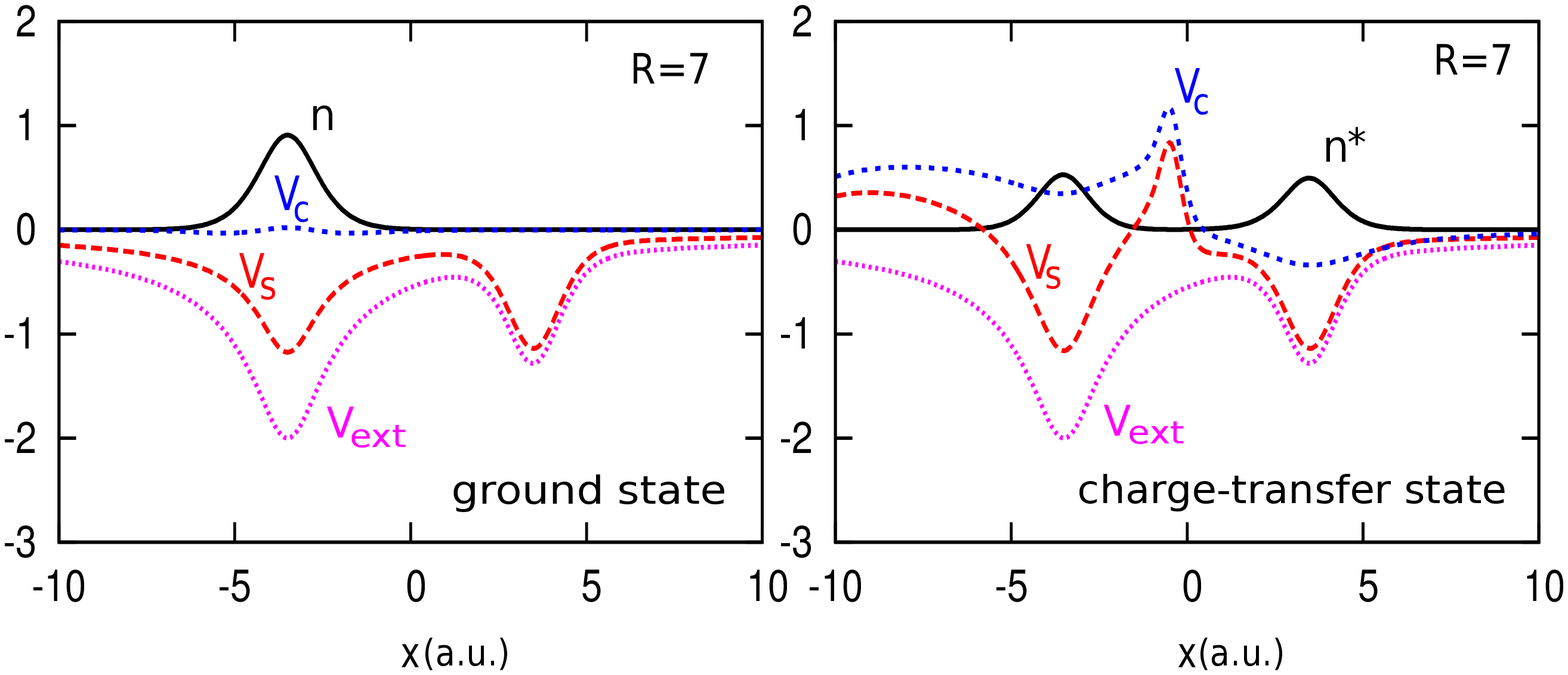}
%R7_box_30.eps
\end{center}
\caption{Density (black solid), $v\s$ (red long-dashed), $v\c$ (blue dashed), and $v\ext$ (pink dotted) for the ground-state (left) and for the CT state (right) in our model molecule of closed-shell fragments at separation $R=7$au.}
\label{f:R7_den_xc}
\end{figure}

To illustrate the mechanism of CT processes and the relevance of spatial and time non-locality 
we use a ``two-electron molecule'' in one-dimension. The Hamiltonian is (atomic units are used throughout):
\bea
H(x_1,x_2,t) &=& -\half\frac{\partial^2}{\partial x_1^2}-\half\frac{\partial^2}{\partial x_2^2} + v\mol(x_1) + v\mol(x_2) \nonumber \\
& &+ v\ee(x_1-x_2) + {\cal E}(t)\cdot(x_1+x_2)
\eea
where $v\ee(y) =1/\sqrt{y^2+1}$ is the ``soft-Coulomb'' electron-electron interaction~\cite{JES88, LSWBKGE96,VIC96,BN02,BL05, K01, KLG04}, and ${\cal E}(t)=A \cos(\omega t)$ is an applied electric field. The molecule is modeled by:
\ben
v\mol(x) = 
%-Z/\sqrt{(x + R/2)^2 + 1} -U_0/\cosh^2(x - R/2)
\frac{-Z}{\sqrt{(x + \frac{R}{2})^2 + a}} -\frac{U_0}{\cosh^2(x - \frac{R}{2})}
\label{vmol}
\een Asymptotically the soft-coulomb potential (donor) on the left
decays as $-Z/x$, similar to a true atomic potential in 3D, while the
cosh-squared (acceptor) on the right is short-ranged, decaying
exponentially away from the ``atom''.  The acceptor potential mimics a
closed-shell atom without  core electrons. 
%By varying the parameters $Z$ and $U_0$, the character of the ground and excited states change. 
We model CT between two closed-shell
fragments, by choosing $Z=2$ and $U_0=1$ such that, at large separations
$R$, the ground-state has two electrons on the donor and zero on the
acceptor, while the first singlet excited state, $\Psi^*$, is a CT excited state
with one electron in each well (see Fig. ~\ref{f:R7_den_xc}). 
%The applied electric field ${\cal E}(t)$ induces this charge transfer.  
%For this case we choose  $Z=2$ and $U_0=1$.
Choosing $Z=2, U_0=1.5$ places one electron in each well in the ground-state, with a
CT excited state having both electrons in the acceptor well;
such a system would model CT between two
open-shell fragments.

%Choosing $Z=2$ and $U=1$, in the infinite-separation limit, the
% ground-state is that of soft-coulomb Helium (E=-2.238 H), and the
% first singlet excited state, $\Psi^*$, is a charge transfer state with one
% electron having moved to the acceptor well. 
%In Fig. \ref{f:ctEvsR}, we
% show how the eigenvalues of this excitation clearly displays the
% characteristic $-1/R$ behavior of a CT states.
 
%We extract the exact xc potential of TDDFT at time $t$ by the
%following method.  We first solve the exact two-electron Schr\"odinger
%equation (e.g. using octopus~\cite{octopus,octopus2}), to obtain the exact
%density $n(x,t)$ and current $j(x,t)$ of the interacting system.  
If we start the KS simulation in a doubly-occupied singlet state, the KS evolution
retains this form for all later times, $\Phi(x_1,x_2,t) = \phi(x_1,t)\phi(x_2,t)$. Requiring
the exact density to be reproduced at all times leads to
%. Since this orbital
%must reproduce the exact density, $n(x,t)$, at all times, the
%magnitude is then fixed. Any phase the orbital may have is then fixed
%by requiring the continuity equation is satisfied, leading to:
%\ben
$
\phi(x,t) = \sqrt{n(x,t)/2}e^{i \int^x dx' u(x',t)}
$
%\een
,where $u(x,t)=j(x,t)/n(x,t)$ is the local ``velocity''. Inverting the KS equation yields the exact KS potential as:
\ben
\label{vsexact}
v\s(x,t)= \frac{\partial^2_x n(x,t)}{4 n(x,t)}-\frac{(\partial_x n(x,t))^2}{8n^2(x,t)} - \frac{ u^2(x,t)}{2} - \int^x\partial_t u(x',t)dx'
\een 
The xc potential is then 
\ben
\label{vxcdef}
v\xc(x,t) = v\s(x,t) - v\ext(x,t) -v\H(x,t)
\een
where $v\H(x,t)=\int dx' n(x',t)v\ee(x-x')$ is the Hartree potential and the external field is given by $v\ext(x,t) = v\mol(x) + {\cal E}(t)x$. Further, for this case, $v\c = v\xc - v\x$, may easily be isolated since $v\x = -v\H/2$.

Before discussing the  dynamics, we first
consider the final CT state, and focus on CT between closed-shell fragments.  Let us assume we have
complete transfer of an electron at some time $T$ into the excited
state $\Psi^*$ (for example applying a tailored laser pulse), and the system then stays in this state for all
times $t>T$.  The density, $n(t>T) = n^*$, is then static in the
excited state and node-less, and the current and velocity $u(x,t)$ are
zero. It follows that the exact $v\xc(t>T)$ is static and that
the exact KS potential is given by first two terms of
Eq. (\ref{vsexact}) only.

In Fig.~\ref{f:R7_den_xc}, we show the density and the exact KS and correlation potentials for the ground and CT states for $R=7$au. 
A clear step and peak
structure has developed in the correlation potential in the region of low-density between the
ions in the CT state. There is no such structure in the initial potential of the ground-state. As the separation increases, the step in $v\c$ saturates to a size
\ben
%{\rm Step \; in\; } v\c
\Delta = \vert I_D^{N_D-1}-I_A^{N_A+1}\vert\;,
\label{eq:stepvxc}
\een
where $I_D^{N_D-1} =I_D^{N=1}$ is the ionization energy of the donor
containing one electron, $I_A^{N_A+1}=I_A^{N=1}$ is that of the 
one-electron acceptor ion, and  the result is written for a general $N_D(N_A)$-electron donor(acceptor). 
Eq.~(\ref{eq:stepvxc}) can 
%shown by inserting $\phi(x,t>T) = \sqrt{n^*/2} = \sqrt{(\phi_A^2 +\phi_D^2)/2}$
%into the KS equation, where $\phi_{A,D}$ are the lowest orbitals on each
%separated well when one electron occupies each, and requiring reduction to the atomic KS equations near each atom.
%{\bf or simplify as peter had, just:
be shown by considering the asymptotics of the donor and acceptor
orbitals, adapting the argument
made for the case of the {\it ground-state} of a molecule made of {\it open-shell}
fragments~\cite{HTR09,TMM09}. 
Here instead, we have a step in the potential of a CT  excited-state of a molecule made of {\it closed-shell} fragments. 
%The entire step is contained in the correlation potential, as can be seen directly for the two-electron case where $v\x =-v\H/2$. 

Somewhat of the reverse picture occurs for the case of CT between open-shells: the initial ground-state correlation potential contains a
step and peak, as shown earlier in~\cite{P85b,AB85,GB96,TMM09,HTR09}, that disappears as the CT state is reached.  In either case, the step is a signature of 
of the strong correlation due to the delocalization of the KS orbital.

%however with the obvious, but important
%difference, that this is not a ground-state. 
%Furthermore, we can
%deduce the magnitude of the step in the infinite-separation limit to
%be $I_D^{N=1}-I_A^{N=1}$ (where we label the donor well as the donor
%and the acceptor well as the acceptor) by considering the asymptotic decayaaaaa
%of the donor and acceptor $1-e$ orbitals {\bf is this right?}.

%As in the latter ground-state case,  the step in the correlation potential of the CT excited state
%appears for similar reasons as the step
%so-called static correlation step {\it i dont know if it is called this by anyone but us} 
%in ground-state DFT, and it
% and puts in evidence 
The step requires a spatially non-local density-dependence in the correlation functional, as in the ground-state case \cite{P85b,AB85,GB96,TMM09,HTR09}.
The inability of usual ground-state approximate functionals 
to capture this step results in them incorrectly predicting 
fractionally charged species. 
%\cite{P85bAB85,GB96,TMM09,HTR09}. 
%{\it do all these references study fractional charged species?}
%Neepa: I changed the wording above to closer to what we had before. You had ``This step...puts in evidence the inability of usual ground-state functionals to decribe fractionally charged species''  But there are no fractional charges here an I am not sure what you meant here?
In the present case, we have an excited state of the interacting system, where the KS orbital corresponding to the excited-state density $n^*$ shown in Fig.~\ref{f:R7_den_xc} is in fact a
ground-state orbital, $\phi(x) = \sqrt{n^*(x)/2}$, because $n^*$ has
no nodes.  Given the static ground-state nature of the orbital and
KS potentials after time $T$, does the adiabatic 
approximation become exact?

\begin{figure}[ht]
\begin{center}
\includegraphics[width=0.475\textwidth,height=0.2\textwidth,clip]{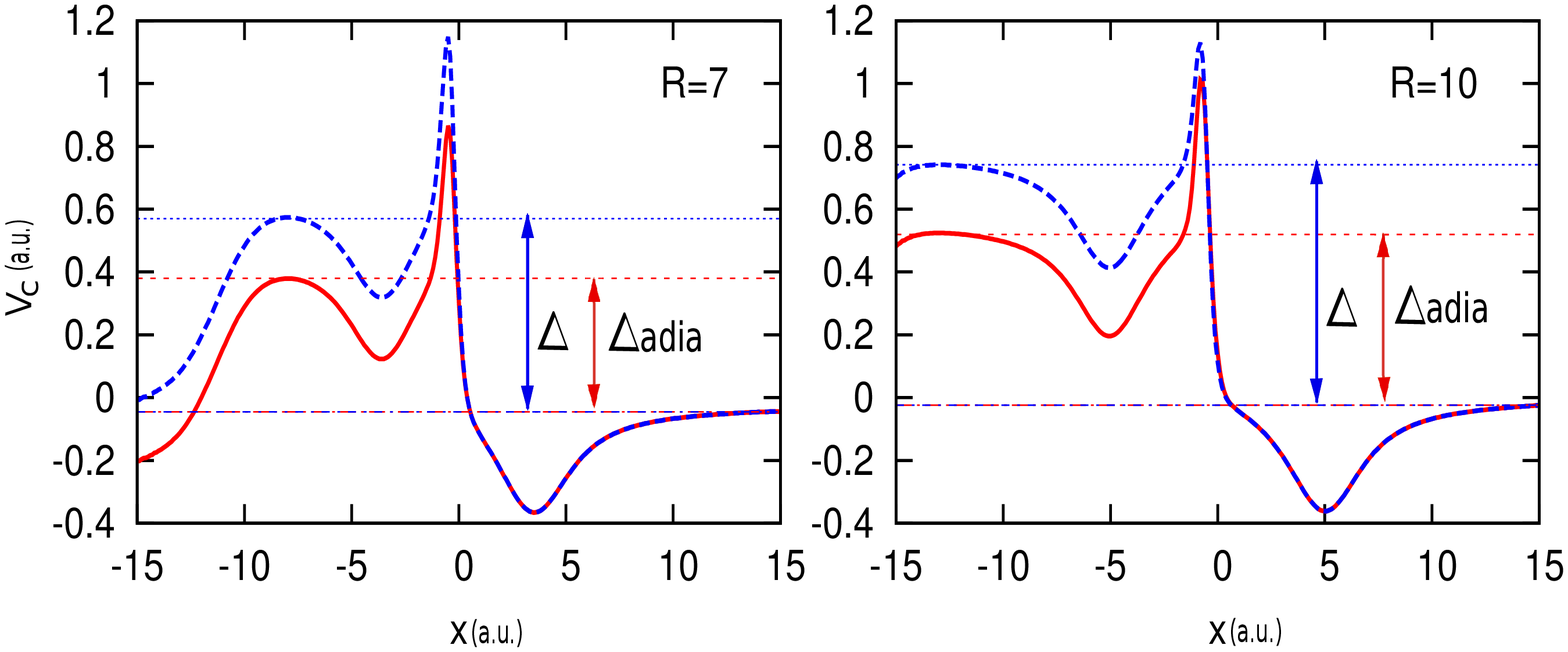}
\end{center}
\caption{The exact $v\c$ (dashed blue line) and the adiabatically-exact $v\c^{adia-ex}$ (red solid line) for  $R=7$au (left) and for $R=10$ au(right). 
Note that the potential eventually rolls back down to zero far enough away from the system. In the infinite separation limit $\Delta$($\Delta_{adia}$)is given by Eq.~(\ref{eq:stepvxc})(Eq.~(\ref{eq:stepxcadia})).
}
%{\bf (i) is this really vc? it looks more like vxc}.{\bf(ii) kink on the right green at R=10 needs explanation, otherwise just put R=7, i guess the R=10 doesn't really add that much  to the figure anyway, so maybe better just to put R=7}}
\label{f:R7_ct_vxc_adia}
\end{figure}

To answer this, we examine the {\it adiabatically-exact} xc potential
for $t>T$, $v\xc^{adia-ex}[\n^*]$, i.e. evaluating the exact
ground-state xc functional on the instantaneous CT density. This is (see Refs.~\cite{TGK08,EM12}):
\ben
\label{vxcadia}
v\xc^{adia-ex}[\n] = v\s^{adia}[\n]-v\ext^{adia}[\n]-v\H[n]
\een
where $v\ext^{adia}[\n]$($v\s^{adia}[\n]$ ) is the external(exact ground-state KS) potential for two
interacting electrons in a ground-state of this density ($v\s^{adia}[\n]$ corresponds to first two terms of Eq.~(\ref{vsexact})).
% and $v\s^{adia}[\n]$ is the exact ground-state KS potential for this density (first two terms of Eq.~(\ref{vsexact})).
Fig.~\ref{f:R7_ct_vxc_adia} shows $v\c^{adia-ex}[\n^*]$ for two
separations $R=7$au and 10au (see Supporting Information for numerical methods). 
%found using similar numerical techniques to
%Refs.~\onlinecite{TGK08} and \onlinecite{EM12}.  
Evidently, the
adiabatic approximation does yield a step, but of the wrong size.

To understand this, first consider the functional dependence
of the exact xc potential. We may write~(\cite{MBW02})
\ben
v\xc[n](t>T) = v\xc[n^*,\Psi^*,\Phi_{CT}^{gs}](t>T)\;,
\label{eq:exactxc}
\een
where, on the left, the dependence is on the entire history of the
density, $n(0<t<T)$, and initial-state dependence is not needed since
at $t=0$ we start from 
%the initial time the interacting and KS states are at 
the ground-state~\cite{newTDDFTbook,MBW02}. On the right,
%history-dependence has been ``traded'' for initial-state
%dependence; 
time $T$ is considered as the ``initial''
time, and the functional depends on just the static density $n^*$
after this time, but, crucially, the interacting state and KS states
at time $T$. The former is the CT excited state $\Psi^*$,
while the latter is the doubly-occupied orbital: $\Phi(x_1,x_2,T) =
\sqrt{n^*(x_1)n^*(x_2)}/2 \equiv \Phi_{CT}^{gs}$, a ground-state
wavefunction, as discussed above.

On the other hand, the adiabatic approximation
\ben
v\xc^{adia}[n^*] \equiv v\xc^{adia}[n^*,\Psi_{CT}^{gs},\Phi_{CT}^{gs}]\;,
\label{eq:adiaxc}
\een
differs from the exact xc potential Eq.~(\ref{eq:exactxc}), in its dependence on the time-$T$ interacting state: here $\Psi_{CT}^{gs}$ is the {\it ground-state} wavefunction of an interacting system with density $n^*$, {\it not the true excited state wavefunction}. 
%As this point, we should bring up the possibility that this potential may not exist due to the density not being {\it v-representable} in either the interacting or non-interacting systems. However for now we will continue with this analysis with the assumption is does exist.
Therefore, Eqs.~(\ref{eq:exactxc}) and~(\ref{eq:adiaxc}) show that the
adiabatically-exact xc potential is not the same as the exact xc
potential: the initial-state dependence in the exact functional reflects a nonlocal time-dependence that persists forever. 
In the infinite-separation limit, we expect $\Psi^*$ and
$\Psi_{CT}^{gs}$ to be very similar, both having a Heitler-London
form with one electron in each well, but the fact that $\Psi^*$ is an excited state is encoded
in the nodal structure of its wavefunction. The correlation potential is extremely sensitive to this tiny difference in the two interacting wavefunctions, which accounts for the different step size in Fig.~\ref{f:R7_ct_vxc_adia}.

The magnitude of the step in $v\xc^{adia-ex}$ in the
infinite-separation limit can be derived by examining the terms in
Eq.~(\ref{vxcadia}).
%\ben
%\Psi_{CT}^*(R \rightarrow \infty)= \frac{1}{\sqrt{2}}(\phi_L(x_1) \phi_R(x_2) +
%\phi_L(x_2)\phi_R(x_1))
%\een
In this limit, locally around each well $v\ext^{adia}$ must equal the
atomic potential, up to a spatial constant, in order for  $\Psi_{CT}^{gs}[n^*]$ to satisfy Schr\"odinger's
equation there. It cannot simply be the sum of the atomic potentials,
because the ground-state $\Psi_0$ of that potential (Eq.~\ref{vmol}) places two
electrons in the donor well.  For $\Psi_{CT}^{gs}$ to be the
ground state, $v\ext^{adia}$ has a step in the region of negligible density that
pushes up the donor well relative to the acceptor well; the size of this step, $C$, is the lowest such that energetically it is favorable to place one electron on each well, as $\Psi_{CT}^{gs}[n^*]$ does.  So, 
%$\Psi_{CT}^{gs}$, the groundhe ground-state wavefunction,-state wavefunction of $v\ext^{adia}[\n^*]$, will be of HL form as discussed. In order to make $\Psi_{CT}^{gs}$ the ground-state wavefunction,
%we imagine adding a step function with height $C$ to the atomic potentials, i.e only the donor well (the donor) is pushed up by this step. At some value of $C$, the soft-coulomb Helium ground-state is no longer lower in energy than the CT case, and it becomes the new ground state. The condition on $C$ is thus
\ben
E^{gs, N=1}_D + E_A^{gs,N=1} +C < E^{gs,N=2}_D + 2C
\een
where $E^{gs,N}_{D(A)}$ is the ground-state energy of the N-electron donor(acceptor). 
%where the left-hand-side represents the energy  $E[\Psi_{CT}^{gs}]$ and the
%right-hand-side represents the energy of the lowest state of $v\ext^{adia}$ that has two electrons in the donor well.
This leads to
\bea
C&\ge& E^{gs,N=1}_D + E_{A}^{gs,N=1}  -  E^{gs, N=2}_{D} = I_D^{N_D} - I_A^{N_A+1}
%\\
%&=& A_D^{N=1} - I_A^{N=1} = I_D^{N_D} - I_A^{N_A}
\eea
where in the last line, we have generalized the result to a donor(acceptor) with $N_D(N_A)$ electrons. 
%{\it please check i wrote the correct thing here and in the eqns}
%Due t the energetic cost of the step, in practise $C$ will achieve the minimum it can to ensure this, i.e. the equality. 

Now that we have the step in $v\ext^{adia}[\n^*]$, we
use Eq. (\ref{vxcadia}) to quantify the step in
$v\xc^{adia-ex}[\n^*]$. Since $v\s^{adia} = v\s^{\rm exact}$ here, 
Eq.~\ref{eq:stepvxc} tells us that the step in $v\c^{adia-ex}$ is
%$v\s^{adia}$ is $I_D^{N=1}-I_A^{N=1}$ (as $v\s^{adia} = v\s^{exact}$ in this case). Therefore the step in $v_{xc}^{adiab}$ is,
\ben
%{\rm Step \; in\;}v\c^{adia}
\Delta_{adia} = |I_{D}^{N_D-1} - A_{D}^{N_D-1}|
\label{eq:stepxcadia}
\een
which is equal to the {\it derivative discontinuity} of the $(N_D-1)$-electron donor. 
(As before,  the entire step is contained in the correlation potential).
%\it Notice that since for the two-electron singlet the exchange contribution to the ks potential is equal to $v_H/2$ and since the Hartree potential has no step, the predicted size of the step is the same for xc than for the correlation potential}.
For our system $I_{D}^{N=1}=1.483$au, $A_{D}^{N=1} = 0.755$au and $I_A^{N=1}
= 0.5$au, thus in the infinite separation limit we get a step of
$0.983(0.729)$au in the exact $v\c$($v\c^{adia}$).  The numerical results verify this analysis; the steps
shown in Figure~\ref{f:R7_ct_vxc_adia} for separation $R=7$($R=10$)au have
values of $0.61$ (0.76)au in the exact $v\c$ and $0.42$ (0.55)au in $v\c^{adia}$. For larger separations, the steps tend towards the asymptotic values predicted by the analysis above.
%(we do not plot these due to numerical noise in the low-density region, that does not however appear to alter the net step)
%for the steps in Eq.~\ref{eq:stepvxc} and
%\ref{eq:stepxcadia} respectively and for $R=10$ the size of the step
%in $v_c^{ex}$ yields $0.78$ and in $v_c^{adiab}$ $0.69$.  {\it The
%  numbers are true for $R=10$ only if we measure the size of the step
%  in the $v_c^{adiab}$ this way: the value of the green curve at
%  $x=-20$ minus the value of the red curve at $x=20$ !! !}
%\ben
%v_{xc}^{adiab}[n^*,\Psi_{gs},\phi^s_{gs}]=
%v_{xc}^{ex}[n^*,\Psi_{CT}^*,\phi^s_{gs}]- v_{ext}^{adiab-all} +v_{ext}^{atoms} 
%\label{eq:vxc-adiab}
%\een

%Hence the exact adiabatic xc potential does not have the correct step
%height, even in infinite separation limit. 

In the above analysis, the adiabatically-exact potential was evaluated
on the exact density, as is commonly done when assessing
functionals~\cite{TGK08}, rather than on that obtained from a
self-consistent adiabatic propagation. The latter would likely lead to
an erroneous density at time $T$, but the analysis shows that even
with the exact density at time $T$, the wrong step-size means that
subsequent  propagation using the adiabatically-exact
potential will yield the wrong dynamics.

Having studied how the xc potential looks for the final CT state,
we now study how the potential evolves in time to reach such a
state. To simplify the analysis  we exploit Rabi physics to reduce this problem to a
two-state system.  This approach is justified for weak
resonant driving field, and verified numerically by comparing the results with the exact time-dependent wavefunction
found using {\tt octopus} \cite{octopus,octopus2,octopus3}. 
%(see also Ref.\onlinecite{EFRM12}).
The interacting wavefunction may be written as $|\Psi(t)\rangle = a_g(t) |\Psi^{gs}\rangle  + a_e(t) |\Psi^*\rangle$, where 
\ben
i \partial_t  \left( {\begin{array}{c}
 a_g(t)\\
 a_e(t)  \\
 \end{array} } \right)=
\left( {\begin{array}{cc}
 E_g -d_{gg}{\cal E}(t)  &  -d_{eg}{\cal E}(t)  \\
-d_{eg}{\cal E}(t) & E_e -d_{ee}{\cal E}(t)  \\
 \end{array} } \right) \left( {\begin{array}{c}
 a_g(t)   \\
 a_e(t)
 \end{array} } \right)
\label{coeffsEOM}
\een
with $d_{eg}=d_{ge}=0.231$, $d_{gg}=7$ and $d_{ee}=0$ for our system.
%The strength of the field is choosen to be $A=0.006$
The electric field is resonant with the first excitation: ${\cal E}(t) = 0.006\cos(0.112 t)$.

\begin{figure}[t]
 \begin{center}
   \includegraphics[width=0.475\textwidth,height =0.35 \textwidth,clip]{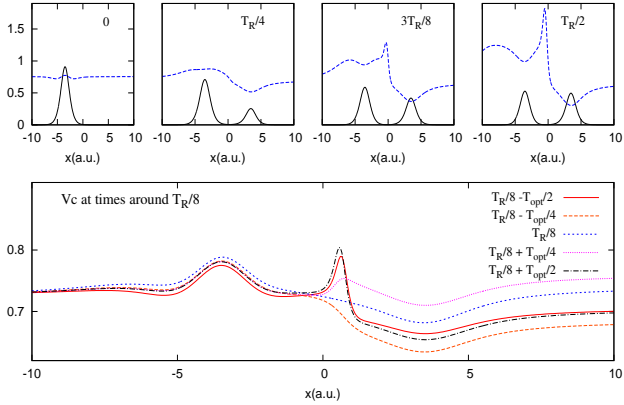}
%Fig3.eps} 
 \end{center}
 \caption{\label{f:CT_vs_snaps}
%{\bf add the T=.. label to each graph; would be great to get the adiabatic vc eg for the last panel. Align the last panel on one side,say the left.} 
Upper panel: The correlation potential (dotted blue line) and density (solid black) shown at snapshots of time indicated. Lower panel: Vc at snapshots over an optical cycle centered around $T_R/8$.  
% are shown in the lower panel.
%\\{\it Fig 3: Add labels: Vc and the label telling at which timepoint each plot is taken.  Angel suggests to get rid of the second plot and instead do the last plot bigger. For t=TR/2 indicate again the size of the step by 2 horizontal lines and the notation $\Delta_{R=7}$.
%The last plot needs explanation.   Angel sugggest to do it arround TR/2 instead of TR/8 (times smaller than TR/2 in the optical scale. Neepa says that around TR/2 dynamical step is very small since acc is very small -- see Figs 32-33 in the notes8-4.pdf file, so Neepa thinks better to stick with TR/8.. ) }
}
\end{figure}

 Fig. \ref{f:CT_vs_snaps} displays the correlation potential at
 snapshots in time over a half-Rabi period $T_R/2$ \cite{fnote2}.
% {\it perhaps we should do this at the weaker field strength 0.006 to make the Rabi two-state soln a better approx to the exact}. 
The step, accompanied by a
 peak, develops over time as the excited CT state is
 reached; at $T_R/2$ the correlation potential agrees
 with the static prediction earlier (Fig.~\ref{f:R7_ct_vxc_adia}, left). 
Notice that making a time-dependent constant shift does not affect the dynamics, just adds a time-dependent overall phase. 
During the second half of the Rabi cycle, the step gradually disappears.
A closer inspection indicates that superimposed to this smoothly developing
 step, is an oscillatory step structure, whose dynamics is more on the
 time-scale of the optical field (lower panel). This faster, non-adiabatic, non-local
 dynamical step appears generically in electron dynamics, as shown in Ref.~\cite{EFRM12}. To distinguish between the two steps we refer to the more gradually developing step due
 to CT, as the ``CT step''.

%For CT between open-shells, somewhat of the
%reverse picture occurs: the initial correlation potential contains a
%step and peak~\cite{P85bAB85,GB96,TMM09,HTR09}, that disappears
%in time as the CT state is reached. 
%When studying
%this process using two electrons however, complications due to
%non-$v$-representability arise~\cite{longerpaper}; at large
%separations one approaches a node in the density in the excited-state,
%leading to a node in the single orbital that is doubly-occupied,
%giving delta-like peaks in the potential. 

%\begin{figure}[t]
% \begin{center}
%  \includegraphics[width=0.2\textwidth,height=0.4\textwidth,angle=270,clip]{vc_%CTZoom.ps} 
% \end{center}
% \caption{\label{f:CT_vc_strobe} The correlation potential and density shown at% snapshots of time indicated around $T_R/8$.{\it merge with previous graph...see there}}
%\end{figure}

The impact that the development of the CT step
has on dynamics is significant. The same adiabatic approximations 
%that show good linear response spectra  and 
that for local resonant excitations showed faster but still Rabi-like
oscillations \cite{FHTR11}, fail dramatically to capture {\it any}
Rabi-like oscillations between the ground and CT state.
%% the failure Common approximations in the adiabatic
%% approximation fail to capture any Rabi-type oscillations between the
%% ground and CT state, 
This is illustrated by the dipole moments, $d(t)=\langle\psi(t)|\hat{x}_1 + \hat{x}_2 |\psi(t)\rangle$, in Figure~({\ref{f:R7}). 
%The dynamical step plays a fundamental role here
The approximate correlation functionals lack the non-local spatial-dependence necessary to develop the CT step
%  potential needed to ``hold'' the electron as it transfers {\bf too vague?}
(\cite{fnote3}).

\begin{figure}
\begin{center}
\includegraphics[width=0.400\textwidth, clip]{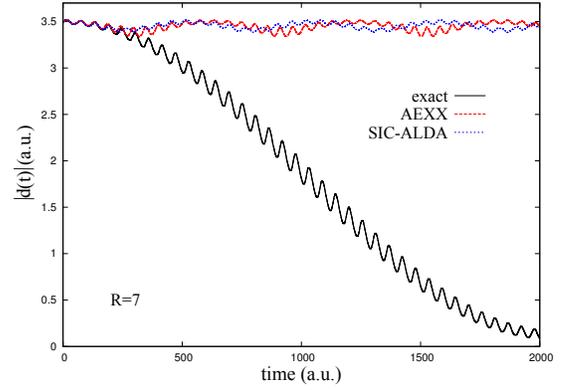}
\end{center}
\caption{
Absolute value dipole moments $|d(t)|$ for the CT between closed-shell fragments at separation $R=7$au for exact (solid black line), adiabatic exact-exchange (AEXX) (dashed red line) and self-interaction-corrected adiabatic local density approximation (SIC-ALDA)(dotted blue line).
The calculations were performed in the presence of a resonant field of frequency $\omega=0.112$ and amplitude $A=0.00667$.
}
\label{f:R7}
\end{figure}

Given the ubiquity of CT dynamics in topical applications
of TDDFT, it is critical to develop approximations with spatially
non-local and non-adiabatic dependence.  None of the available
functionals today captures the peak and step structure that develop in
the exact $v\c$ as the charge transfers, and they lead
to drastically incorrect dynamics, as illustrated in Figure~\ref{f:R7}.
%e.g. dipole oscillations between
%the ground and CT states induced by a weak resonant field
%are completely lacking.  
Even an exact adiabatic approximation will be
incorrect: a step and peak feature are captured but of the wrong
size. The performance of a self-consistent propagation in such a
potential is left for a future investigation, as is the role of the
peak that accompanies the step. Superimposed on the development of the CT step, are the generic dynamical step and peak
features of Ref~\cite{EFRM12}: 
this features depends on the details of how the CT is
induced, e.g. oscillating on the time-scale of a resonant optical
field. Note that the CT step recedes asymptotically
far from the molecule~\cite{HTR09,TMM09}, while the dynamical step persists \cite{EFRM12}.   
The relation of these structures to the derivative discontinuities of the xc kernel for CT excitations~\cite{HG12} will also be investigated in the future.
%Although our studies here involve only two-electron systems, we argue
%that the essential elements will go through for real molecular systems. 

In modeling real systems the vibronic coupling introduces a mixture of
excited states that are not, in principle, fully populated. Still our
findings apply, since for an ensemble of states CT steps and dynamical
steps appear that account for the population of each excited state
contributing to the wave packet.  Note that the step responsible for
the CT appears as soon as the state starts to be populated. Recent
work has shown that TDDFT describes the CT process in an organic
photovoltaic~\cite{Rozzi}; our findings may explain the observed incomplete CT of the electron to the fullerene~\cite{Rozziprivate}. The step feature is a fundamental one for describing processes where electron-hole splitting is key. 
%Our findings may well explain the incomplete CT observed in Ref.~\cite{Rozzi}.
%The present conclusions would be valid for real molecular systems beyond the 2e test case studied here.
%Our work opens the path towards the development of nonadiabatic functionals able to capture 
%(as the potential in Eq.~({\ref{eq:vsexact}})) 
%dynamical electron transfer  processes.
Our work highlights an essential new feature that must be considered in the
development of nonadiabatic functionals able to capture
dynamical electron transfer processes.

{\it Acknowledgments} 
We gratefully acknowledge financial support from the US Department of Energy, Office of Basic Energy Sciences, Division of Chemical Sciences, Geosciences and Biosciences and Award DE-SC0008623 (NTM), and
a grant of computer time from the CUNY High Performance
Computing Center under NSF Grants CNS-0855217. 
%The European Research Council (ERC-2010-
%AdG -Proposal No. 267374) Spanish Grants (FIS2011-
%65702- C02-01 and PIB2010US-00652), Grupos Consolidados
%(IT-319-07), and European Commission project
%CRONOS (280879-2) and CNS-0958379, are gratefully
%acknowledged. JIF acknowledges support from an FPI fellowship
%(FIS2007-65702-C02-01).
The European Research Council (ERC-2010- AdG -267374) , Spanish: FIS2011- 65702- C02-01), Grupos Consolidados (IT-319-07), and EC project CRONOS (280879-2) and CNS- 0958379, are gratefully acknowledged. JIF acknowledges support from an FPI fellowship (FIS2007-65702- C02-01). 
%\end{acknowledgments}

%\begin{sectio}
Supporting Information Available: All numerical aspects of the calculations are contained in the Supporting Information. 
This information is available free of charge via the Internet at http://pubs.acs.org/.
%\end{section}


\begin{thebibliography}{99}

%\bibitem{T12}
%{\em Perspective: Nonadiabatic dynamics theory},
%J. C. Tully , J. Chem. Phys. {\bf 137}, 22A301 (2012).

\bibitem{RG84}
Runge, E.; Gross, E.K.U. 
 Density-Functional Theory for Time-Dependent Systems.
{\it Phys. Rev. Lett.} {\bf 1984}, {\it 52}, 997-1000.

\bibitem{newTDDFTbook}
{\it Fundamentals of Time-Dependent Density Functional Theory, (Lecture Notes in Physics 837)},   eds. Marques, M.A. L.; Maitra, N.T.; Nogueira, F.;  Gross, E.K.U.; Rubio, A. (Springer-Verlag, Berlin, Heidelberg, 2012).         

\bibitem{PCCPissue}
Themed Issue on Time-Dependent Density Functional Theory. {\it Phys. Chem. Chem. Phys} {\bf 2009}, {\it 11}, eds. Marques, M. A. L.; Rubio, A. 

\bibitem{A09}
Autschbach, J., Charge-Transfer Excitations and Time-Dependent Density
Functional Theory: Problems and Some Proposed
Solutions. {\it ChemPhysChem} {\bf 2009}, {\it 10}, 1757-1760, and references therein.

%\bibitem{SKB09}
%T. Stein, L. Kronik, and R. Baer,
%Reliable Prediction of Charge Transfer Excitations in Molecular Complexes Using Time-Dependent Density Functional Theory, 
%J. Am. Chem. Soc. {\bf 131}, 2818 (2009).

\bibitem{SKB09}
Stein, T.; Kronik, L.; Baer, R. Reliable Prediction of Charge Transfer Excitations in Molecular Complexes
Using Time-Dependent Density Functional Theory. {\it J. Am. Chem. Soc.} {\bf 2009}, {\it 131}, 2818--2821.

\bibitem{BLS10}
Baer, R.; Livshits, E.; Salzner, U. Tuned Range-Separated
Hybrids in Density Functional
Theory. {\it Annu. Rev. Phys. Chem.} {\bf 2010}, {\it 61}, 85--109. 
%Tuned Range-Separated Hybrids in Density Functional Theory

\bibitem{HG12}
Hellgren, M.; Gross, E. K. U., Discontinuities of the Exchange-Correlation Kernel and Charge-Transfer Excitations
in Time-Dependent Density-Functional Theory. {\it Phys. Rev. A.} {\bf 2012}, {\it 85}, 022514-1--6.

\bibitem{M05c}
Maitra, N.\,T. Undoing Static Correlation: Long-Range Charge Transfer in Time-Dependent Density-Functional Theory, 
{\em J. Chem. Phys.} {\bf 2005}, {\it 122}, 234104-1--6.

\bibitem{MT06}
Maitra, N. T.; Tempel, D. G. Long-Range Excitations in Time-Dependent Density-Functional Theory. {\em J. Chem. Phys.} {\bf 2006}, {\it 125}, 184111-1--6.

\bibitem{FRM11}
Fuks, J. I.; Rubio, A.; Maitra, N.T. Charge Transfer in Time-Dependent Density-Functional Theory via Spin-Symmetry Breaking, {\it Phys. Rev. A.} {\bf 2011}, {\it 83}, 042501-1--5.

\bibitem{TMM09}
Tempel, D. G.; Maitra, N. T.; Martinez, T. J. Revisiting Molecular Dissociation in
Density Functional Theory: A Simple Model. {\it J. Chem. Theory Comput.} {\bf 2009}, {\it 5}, 770--780.

\bibitem{HTR09}
Helbig, N.; Tokatly, I. V.; Rubio, A. Exact Kohn-Sham Potential of Strongly Correlated Finite Systems. {\it J. Chem. Phys.} {\bf 2009}, {\it 131}, 224105-1--8. 

\bibitem{PDP09}
Prezhdo, O. V.; W. R. Duncan, W. R.; and Prezhdo, V. V. Photoinduced Electron Dynamics at the Chromophore-Semiconductor Interface: A Time-Domain Ab Initio Perspective. {\it Prog. Surf. Sci.} {\bf 2009}, {\it 84}, 30--68. 

\bibitem{Oviedo}
Oviedo, M.; Zarate, X.; Negre, C. F. A.; Schott, E.; Arriatia-P\'erez R.; S\'anchez, C. G. Quantum Dynamical Simulations as a Tool for Predicting Photoinjection Mechanisms in Dye-Sensitized Ti02 Solar Cells. {\it J. Phys. Chem. Lett.} {\bf 2012}, {\it 3}, 2548--2555.

\bibitem{Teoh}
Teoh, W. Y.; Scott, J. A.; Amal, R. Progress in Heterogeneous Photocatalysis: From Classical Radical Chemistry to Engineering Nanomaterials and Solar Reactors. {\it J. Phys. Chem. Lett.} {\bf 2012}, {\it 3}, 629--639. 

\bibitem{Nyuen}
Nguyen, P. et al. Solvated First-Principles Excited-State Charge-Transfer Dynamics with Time-Dependent Polarizable Continuum Model and Solvent Dielectric Relaxation. {\it J. Phys. Chem. Lett.} {\bf 2012}, {\it 3}, 2898--2904.

\bibitem{Moore}
Moore, J.E.; Morton, S.M.; Jensen, L. Importance of Correctly Describing Charge-Transfer Excitations for Understanding the Chemical Effect in SERS. {\it J. Phys. Chem. Lett.} {\bf 2012}, {\it 3}, 2470--2475

\bibitem{Meng}
Meng, S.; Kaxiras,E. Electron and Hole Dynamics in Dye-Sensitized Solar Cells: Influencing Factors and Systematic Trends. {\it Nano Lett.} {\bf 2010}, {\it 10}, 1238--1247. 

\bibitem{fnote1}
Some degree of spatial and time non-locality is introduced in orbital functionals, such as hybrid functionals that mix in a fraction of non-local exchange~\cite{KK09,BLS10}.

\bibitem{KK09}
K\"ummel, S.; Kronik, L. Orbital-Dependent Density Functionals: Theory and Applications. {\it Rev. Mod. Phys.} {\bf 2008}, {\it 80}, 3--60.


\bibitem{EFRM12}
Elliot, P.; Fuks, J. I.; Maitra, N.T.; Rubio, A. Universal Dynamical Steps in the Exact Time-Dependent Exchange-Correlation Potential. {\it Phys. Rev. Lett.} {\bf 2012}, {\it 109}, 266404-1--5.

\bibitem{JES88}
Javanainen, J.; Eberly, J.; Su, Q. 
Numerical simulations of multiphoton ionization and above-threshold electron spectra. {\it Phys. Rev. A.} {\bf 1988}, {\it 38}, 3430--3446.

\bibitem{LSWBKGE96}
Lappas, D.G.; Sanpera, A.; Watson, J.B.; Burnett, K.; Knight, P.L; Grobe, R.; Eberly, J.H. Two-electron effects in harmonic generation and ionization from a model He atom. {\it J. Phys. B}, {\bf 1996}, {\it 29},  L619 , 16

\bibitem{VIC96}
Villeneuve, D.M.; Ivanov, M.Y.;Corkum, P.B.
Enhanced ionization of diatomic molecules in strong laser fields:
A classical model.
{\it Phys.Rev. A.} {\bf 1996}, {\it 54}, 736--741

\bibitem{BN02}
Bandrauk, A.;  Ngyuen, H.
Attosecond control of ionization and high-order harmonic generation
in molecules.
{\it Phys. Rev. A.} {\bf 2002}, {\it 66}, 031401

\bibitem{BL05} 
Bandrauk, A.D.; Lu, H.
Laser-induced electron recollision in H2 and electron correlation.
{\it Phys. Rev. A}{\bf 2005},{\it 72}, 023408

\bibitem{K01}
Kreibich , T.  {\it et~al.}.
Even-Harmonic Generation due to Beyond-Born-Oppenheimer Dynamics.
{\it Phys. Rev. Lett.} {\bf 2001},{\it 87}, 103901

\bibitem{KLG04}
Kreibich, T.; van Leeuwen, R.; Gross, E. K. U..
Time-dependent variational approach to molecules in strong laser fields. {\it Chem. Phys.}{\bf 2004}, {\it 304}, 183-202.

%{JES88,LSWB96,VIC96,BN02,KLG04}VIC96,LL98,LKGE02,KLG04,KLEG01,BL05,BN02,CCZB96}
%BS02, 
%JES88, VIC96, L96, BS02, KLG04


%\bibitem{LSWB96}
%D.G. Lappas, A. Sanpera, J.B. Watson, K. Burnett, P.L. Knight, R. Grobe, J.H. Eberly, J. Phys. B 29,  L619 (1996). 

%\bibitem{BN02}
% A. Bandrauk and H. Ngyuen, Phys. Rev. A. {\it 66}, 031401(R) (2002). 

%\bibitem{BL05}
%A. D. Bandrauk and H. Lu, Phys. Rev. A. {\it 72}, 023408 (2005). 

%\bibitem{KLEG01}
%T. Kreibich et al. Phys, Rev. Lett. {\it 87}, 103901 (2001).

%\bibitem{KLG04}

%\bibitem{VIC96}
%D.M. Villeneuve, M.Y. Ivanov, and P.B. Corkum, Phys.Rev. A. {\it 54}, 736 (1996).

\bibitem{P85b}
Perdew, J. P. in {\it Density Functional Methods in Physics}, edited by 
Dreizler R.M. and da Providencia, J.;  Plenum: New York, 1985.

\bibitem{AB85}
Almbladh, C. O. ;von Barth, U. in {\it Density Functional Methods in Physics}, edited by 
Dreizler, R.M. and da Providencia, J.;  Plenum: New York, 1985.

%\bibitem{AB85}
%C. O. Almbladh and U. von Barth, in {\it Density Functional Methods in
%Physics}, edited by R. M. Dreizler and J. da Providencia, Plenum, New
%York, 1985.

\bibitem{GB96}
Gritsenko, O.V.; Baerends, E.J. Effect of Molecular Dissociation on the Exchange-Correlation Kohn-Sham Potential. {\it Phys. Rev. A} {\bf 1996}, {\it 54}, 1957 --1972.


\bibitem{octopus}
Castro, A. {\it et~al.}. Octopus: A Tool for the Application of Time-Dependent Density
Functional Theory. {\it Phys. Stat. Sol. (b)} {\bf 2006}, {\it 243}, 2465--2488.

\bibitem{octopus2}
Marques, M. A. L.; Castro, A.; Bertsch, G. F.; Rubio, A. Octopus: A First-Principles Tool
for Excited Electron-Ion Dynamics. {\it Comp. Phys. Comm.} {\bf 2003}, {\it 151}, 60--78.

\bibitem{octopus3}
Andrade, X. {\it et al}. Time-Dependent Density-Functional Theory in Massively Parallel Computer Architectures: The Octopus Project. {\it J. Phys.: Condens. Matter} {\bf 2012}, {\it 24} 233202--1--11.


\bibitem{TGK08}
Thiele, M.; Gross, E.K.U.; K\"ummel, S. Adiabatic Approximation in Nonperturbative Time-Dependent Density-Functional Theory. {\it Phys. Rev. Lett.} {\bf 2008}, {\it 100}, 153004-1--4.

\bibitem{EM12}
Elliott, P.; Maitra, N. T. Propagation of Initially Excited States in Time-Dependent Density Functional Theory. {\it Phys. Rev. A.} {\bf 2012}, {\it 85}, 052510-1--11.

\bibitem{MBW02}
Maitra, N.T.; Burke, K.; Woodward, C. Memory in time-dependent density functional theory. {\it Phys. Rev. Lett.} {\bf 2002},  {\it 89}, 023002-1--4.

%\bibitem{longerpaper}
%J. I. Fuks, P. Elliott, N.T. Maitra, and A. Rubio, in preparation.

\bibitem{fnote2}
Within the rotating wave approximation $T_R=2\pi/(\frac{2 d_{ge}A}{z}J_1[z])$ , where $J_1[z]$ is a Bessel function with argument $z=\frac{(d_{ee}-d_{gg})A}{\omega}$
%, $A$ and $\omega$ are the parameters of the external field and $d_{ge},d_{gg}, d_{ee}$ are matrix elements that characterize the system 
(see Ref.~\cite{BMT00} for details). The CT state for our model is reached at $T_R/2$.

\bibitem{FHTR11}
Fuks, J. I.; Helbig, N.; Tokatly, I.V.; Rubio, A. Nonlinear phenomena in time-dependent density-functional theory: What Rabi oscillations can teach us.
{\it Phys. Rev. B.} {\bf 2011}, {\it 84}, 075107. 

\bibitem{fnote3} For smaller separations the approximations perform better, as their
  character becomes more local; but errors
  remain without non-adiabatic non-local dependence needed in the
  potential~\cite{EFRM12}.

%\bibitem{KM85}
%M. A. Kmetic and W. J. Meath, Phys. Lett. A {\it 108}, 340 (1985).

\bibitem{BMT00}
Brown, A.; Meath, W.J.; Tran, P. Rotating-Wave Approximation for the Interaction of a Pulsed Laser with a Two-Level System Possessing Permanent Dipole Moments. {\it Phys Rev. A.} {\bf 2000}, {\it 63}, 013403-1--7.
.
\bibitem{Rozzi}
Rozzi, C.A. {\it et~al.}
Quantum Coherence Controls the Charge Separation in a Prototypical Artificial Light Harvesting System. Nature Communications {\bf 2013}, in press. 

\bibitem{Rozziprivate}
Rozzi, C.A.; Rubio, A. private communication. 


\end{thebibliography}
\end{document}

\title{Supporting information for: Dynamics of Charge-Transfer Processes with Time-Dependent Density Functional Theory}

\author{J. I. Fuks}
\affiliation{Nano-Bio Spectroscopy group, Dpto.~F\'isica de Materiales, Universidad del Pa\'is Vasco, Centro de
F\'isica de Materiales CSIC-UPV/EHU-MPC and DIPC, Av.~Tolosa 72, E-20018 San 
Sebasti\'an, Spain}
\author{P. Elliott}
\affiliation{Max-Planck-Institut f\"{u}r Mikrostrukturphysik, Weinberg 2, 06120 Halle (Saale), Germany}
\author{A. Rubio}
\affiliation{Nano-Bio Spectroscopy group, Dpto.~F\'isica de Materiales, Universidad del Pa\'is Vasco, Centro de
F\'isica de Materiales CSIC-UPV/EHU-MPC and DIPC, Av.~Tolosa 72, E-20018 San 
Sebasti\'an, Spain}
\affiliation{Fritz-Haber-Institut der Max-Planck-Gesellschaft, Faradayweg 4-6,
D-14195 Berlin, Germany}
\author{N. T. Maitra}
\affiliation{Department of Physics and Astronomy, Hunter College and the City University of New York, 695 Park Avenue, New York, New York 10065, USA}
\date{\today}
\pacs{}

\begin{abstract}
In this supporting material we give additional numerical details behind the results presented in the paper. In the first part computational details on the ground state calculation and on the real-time propagation are given.  Next the procedure to find the exact Kohn Sham potential Eq.(3) and the adiabatic-exact XC potential Eq.(6) are discussed.  
\end{abstract}

\maketitle

%\section{Computational Details}

In all cases we work in $1$D, which significantly reduces the computational cost, in particular, simple numerical integration techniques can be used. We also use a regular real-space grid with spacing at most $0.1$a.u., the entire simulation is then contained within a box of total length $100$a.u. 
%(absorbing boundary conditions are employed for time-dependent runs). 
%\subsection{Ground-State Calculations}
\\The ground-state Hamiltonian, Eq. (1) with ${\cal E}(t)=0$, consists of two interacting electrons in the molecular potential, this can be solved exactly using the open-source electronic structure program {\tt OCTOPUS} \cite{octopus, octopus2, octopus3} by considering the problem as a single electron in a complicated $2D$ potential \cite{HFTAGR11}. By this method, we can find the exact energies and wavefunctions for the ground-state and charge-transfer state. 
Imaginary time propagation using the time-dependent methods discussed below was also used to find the exact ground-state, Gram-Schmidt orthogonalization may be used to find for excited states.
%\subsection{Time-Dependent Calculations}
\\Time-dependent calculations to find the density and current were done by three different methods. The first is, again, {\tt OCTOPUS}, which can perform many different propagation schemes. For these calculations we used approximate enforced time-reversal symmetry or AETRS) with a time-step of $0.01$a.u. The second method is an {\it in-house} code utilizing the exponential mid-point rule to approximate the time-evolution operator and then a forth order expansion of the exponential. A discussion of these techniques can be found in Ref. \onlinecite{castro} . A $9$ point finite-difference rule is used for the laplacian. In this case, a time-step of $0.001$a.u. was used. Both the time propagation methods discussed so far are exact and agree with eachother. The current, $j(x,t)$, was extracted from the wavefunction via standard center-space $2$ point finite-difference for the first derivative. The third and final method is the two-state model of Eq. (12) as introduced in the text. The coefficients $a_g(t)$ and $a_e(
t)$ are found using {\it mathematica}\cite{mathe} and to verify accuracy, also with a coupled $4^{th}$ order Runge-Kutta algorithm. The two-state method is used as the time-derivatives of the coefficients, which are needed to find the current, etc, are given analytically by Eq. (12). It also produces less noise in regions with less density. The calculated densities and currents from the two-state model agree with the exact propagation results up to some small deviations in the low-density tail regions where higher states which are slightly populated by the oscillating field become dominant. This is to be expected and the calculated potentials are not significantly affected in these regions. The self-consistent time-dependent Kohn-Sham calculations shown in Fig. 4 are performed with {\tt OCTOPUS} but with a time-step of $0.025$.
%\subsection{The Kohn-Sham potential}
\\The exact time-dependent Kohn-Sham potential is found from the density and current using Eq. (3). The first two terms are given by applying the $9$-point finite-difference laplacian to the square-root of the density, then dividing by the density. Obviously care must be taken here to avoid numerical noise if the density is too small. Similarly for the local velocity $u(x,t)=j(x,t)/n(x,t)$. Standard numerical integration and differentiation techniques, such as those already discussed are used to calculate the last term of Eq. (3), and for the Hartree potential needed in Eq. (4) to extract the exact XC potential.
%\subsection{Exact adiabatic potentials}
\\The exact-adiabatic XC potential is found using Eq. (6). The only missing piece is to find the external potential for which a given density is the ground-state. We use the method given in Refs. \onlinecite{EM12,TGK08} (and refs. therein) whereby a trial potential is updated based on the difference between the ground-state density and the target density with a prefactor than can be spatial dependent. This is then iterated until convergence. A measure of the difference between the calculated ground-state density for the $k$-th iteration cycle, $n^{(k)}(x)$, and the target density, $n(x)$, 
\begin{equation}
\Delta=\int|n^{(k)}(x)-n(x)|dx
\end{equation}
is calculated as the check for convergence. The imaginary time propagation mentioned above is used to find $n^{(k)}$. Anecdotally we have found that a prefactor of $0.001/\Delta$ converges relatively quickly but this depends heavily on the initial guess for the potential. A weighting based on the inverse of the density was also used to speed up convergence in certain regions. 
\\The charge-transfer problem is particularly demanding as it requires convergence in regions with low density in order to resolve any peak and/or step structure. Often a large peak in $v\ext^{adia}[n]$ will cancel a peak in $v\s^{adia}[n]$ and so care is needed to ensure convergence is reached and the calculated potential is smooth.